\documentclass[journal]{IEEEtran}
\usepackage{amssymb}
\usepackage{amsfonts}
\usepackage{bbm}
\usepackage{amsfonts}
\usepackage[dvips]{graphicx}
\usepackage[dvips]{color}
\usepackage{graphicx}
\usepackage{amsmath}
\usepackage{fancyhdr}
\usepackage{stfloats}
\usepackage{multirow}
\usepackage{algorithm}
\usepackage{algorithmic}
\usepackage{cite}
\usepackage[bookmarks=false]{}
\usepackage{amsthm}
\usepackage{algorithm}
\usepackage{algorithmic}
\usepackage{mathbbol}
\usepackage{amsfonts}
\usepackage{graphicx}
\usepackage{amsmath}
\usepackage{amssymb}
\usepackage{latexsym}
\usepackage{graphicx}
\usepackage{cite}
\usepackage{url}
\usepackage{stfloats}
\usepackage{mathrsfs}
\usepackage{setspace}
\usepackage{booktabs}
\usepackage{latexsym}
\usepackage{url}
\usepackage{stfloats}
\usepackage{mathrsfs}
\theoremstyle{plain}

\usepackage{cite}
\usepackage{setspace}

\usepackage[tight,footnotesize]{subfigure}
\makeatletter
\renewcommand{\maketag@@@}[1]{\hbox{\m@th\normalsize\normalfont#1}}%
\makeatother

\begin{document}
\clearpage
\title{\huge Power Efficient Visible Light Communication (VLC) with Unmanned Aerial Vehicles (UAVs)}
%
\author{\IEEEauthorblockN{Yang Yang\IEEEauthorrefmark{1}, Mingzhe Chen\IEEEauthorrefmark{1}, Caili Guo\IEEEauthorrefmark{1}, Chunyan Feng\IEEEauthorrefmark{1}, and Walid Saad\IEEEauthorrefmark{2}
\\\IEEEauthorrefmark{1}Beijing Key Laboratory of Network System Architecture and Convergence, School of
Information and Communication Engineering, Beijing University of Posts and Telecommunications, Beijing 100876, China
\\\IEEEauthorrefmark{2}Virginia Polytechnic Institute and State University, Blacksburg, USA
\\
}
\vspace{-0.5cm}
\thanks{ This work was supported by National Nature Science Foundation of China (No. 61871047), China Postdoctoral Science Foundation (2018M641278) and the U.S. National Science Foundation under Grant CNS-1836802.
}
}
\vspace{-0.5cm}
%
%
%
%
\vspace{-0.5cm}
\maketitle
\thispagestyle{empty}

\begin{abstract}
A novel approach that combines visible light communication (VLC) with unmanned aerial vehicles (UAVs) to simultaneously provide flexible communication and illumination is proposed.
To minimize the power consumption, the locations of UAVs and the cell associations are optimized under illumination and communication constraints.
An efficient sub-optimal solution that divides the original problem into two sub-problems is proposed.
The first sub-problem is modeled as a classical smallest enclosing disk problem to obtain the optimal locations of UAVs, given the cell association.
Then, assuming fixed UAV locations, the second sub-problem is modeled as a min-size clustering problem to obtain the optimized cell association.
In addition, the obtained UAV locations and cell associations are iteratively optimized multiple times to reduce the power consumption.
Numerical results show that the proposed approach can reduce the total transmit power consumption by at least $53.8\%$  compared to two baseline algorithms with fixed UAV locations.
\end{abstract}
\vspace{-0cm}

\vspace{-0.6cm}
\section{Introduction}
\label{sec:intro}
\IEEEPARstart{V}isible light communication (VLC) relies on light-emitting diodes (LEDs) for signal transmission which makes them particularly suitable for scenarios like search and rescue in which both illumination and communications are required.
Recently, the authors in \cite{Deng2018twinkle} proposed to use, Twinkle, a concept built on the idea of using unmanned aerial vehicles (UAVs) equipped with LEDs, for illumination in applications such as urban safety and disaster recovery. In industry,  Draganfly Innovations is also integrating and testing SureFire LLC’s LED technology with its UAVs for nighttime drone operations.
The use of LED mounted on a UAV motivates us to combine VLC and UAVs to provide both flexible communications and illumination in these applications.


UAVs are often used as aerial base stations or relays to enhance coverage and capacity of wireless networks.
The deployment of UAVs is one of the main challenges in UAV networks.
There is a plethora of prior art on optimal deployment of UAVs \cite{Mozaffari2016Unmanned,Yang2018joint,Chencaching}. The authors in \cite{Mozaffari2016Unmanned} proposed an analytical framework for network coverage and rate analysis under different UAV deployments.
The authors in \cite{Yang2018joint}  minimized uplink power by jointly considering UAVs' altitudes, locations, beamwidth, and bandwidth.
The optimal deployment in terms of quality-of-experience was also investigated for cache-enabled UAVs \cite{Chencaching}.
However, none of these prior studies investigated the joint design of communication and illumination, which is crucial for VLC-enabled UAV networks.

The main contributions of this paper is to propose a novel framework for optimizing the deployment of VLC-equipped UAVs while minimizing power consumption under both illumination and communication constraints. To achieve this goal, we provide the following key contributions:
\vspace{-0.05cm}
\begin{itemize}
\item The effect of illumination and communication constraints on the power of a UAV is analyzed, based on which a power efficient VLC-based UAV network optimization framework is formulated.
\item To address the mutual dependence between the optimization variables, a sub-optimal approach that divides the optimization problem into a UAV location optimization sub-problem and a cell association sub-problem is proposed. These two sub-problems are solved sequentially, and the obtained solutions are iteratively optimized.
\end{itemize}
\vspace{-0.05cm}
Numerical results verify that the proposed scheme can reduce the power consumption by at least $53.8\%$ compared to two baseline algorithms with fixed UAV positions.
To our best knowledge, \emph{this is the first work to jointly consider the use of VLC and UAV to satisfy the users' data rate and illumination requirements}.

\vspace{-0.3cm}
\section{System Model And Problem Formulation}
Consider a geographical area in which downlink communications are provided by $K$ UAVs that communicate with $U$ users. As shown in Fig. 1, UAVs simultaneously provide illumination and broadcast information for a target area. Considering the directional transmission property of VLC, we assume that each user is associated with one UAV.
The channel gain of a VLC link can be represented as: 
\vspace{-0.2cm}
\begin{equation}\label{channel_model}
{h_{ij}} = \left\{ {\begin{array}{*{20}{l}}
{\frac{{(m + 1)A}}{{2\pi d_{ij}^2}}g({\psi _{ij}}){{\cos }^m}({\phi _{ij}})\cos \left( {{\psi _{ij}}} \right),\;0 < {\psi _{ij}} < {\Psi _c}},\\
{\;\;\;\;\;\;\;\;\;\;\;\;\;\;\;\;\;\;\;0,\;\;\;\;\;\;\;\;\;\;\;\;\;\;\;\;\;\;\;\;\;\;\;\;\;\;\;\;\;
\;\psi_{ij}  > {\Psi _c}},
\end{array}} \right.
\end{equation}
\begin{figure}[tp]
\begin{center}
\includegraphics[width=0.5\textwidth]{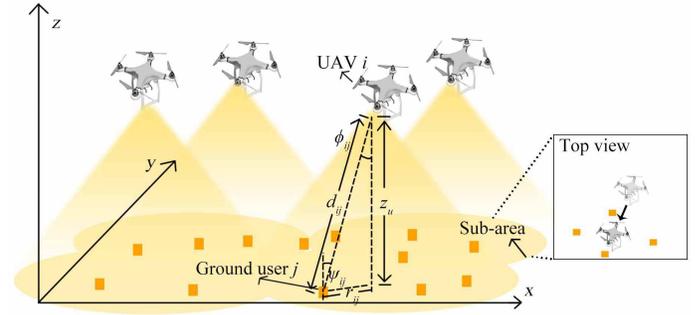}
\vspace{-0.2cm}
\caption{The application scenarios of VLC with UAV.}
\label{scenario}
\end{center}
\end{figure}
where ${h_{ij}}$ denotes the channel gain between UAV $i$ and user $j$, $A$ is the detector area, and $d_{ij}$ is the distance between UAV $i$ and user $j$. In addition, $m =  - \ln 2/\ln (\cos {\Phi _{1/2}})$, where ${\Phi _{1/2}}$ is the transmitter semi-angle, $\psi_{ij}$ is the angle of incidence, $\phi_{ij}$ is the angle of irradiance, ${{\Psi _c}}$ is the receiver field of vision (FOV) semi-angle, and $g(\psi_{ij} )$ is the gain of optical concentrator, which can be given by
$g(\psi_{ij} ) = \left\{ {\begin{array}{*{20}{l}}
{\frac{{{n_r}^2}}{{{{\sin }^2}{\Psi _c}}},\;0 \le {\psi _{ij}} \le {\Psi _c},}\\
{\;\;\;\;0,\;\;\;\;\;{\psi _{ij}} \ge {\Psi _c}},
\end{array}} \right.$
where $n_r$ is the refractive index.
For simplicity, we consider a two-dimensional deployment scenario in which the height of the users is zero while the height of all UAVs is fixed to $z_u$\footnote{Since the value of the channel gain is determined by the height difference between UAVs and users, and not by their absolute values, setting the height of users to zero will not compromise the generality of the derivation process.}.
Given the positions of the UAVs $({x_i},{y_i},z_u)$, $i = 1,2, \ldots K$ and that of the users $({x_j},{y_j},0)$, $j = 1,2, \ldots U$, we have ${d_{ij}} = \sqrt {{{({x_i} - {x_j})}^2} + {{({y_i} - {y_j})}^2} + z_u^2} $ and $\cos {\phi _{ij}} = \cos {\psi _{ij}} = \frac{z_u}{{{d_{ij}}}}$.
Note that these angles are obtained based on the assumption that transmitters and receivers are vertical downward and upward, respectively.
Here, cameras or photodiodes can be deployed as receivers at the ground users so as to convert optical signals into electrical signals and support communications as users move.

\vspace{0.0cm}
The channel capacity $C$ of the VLC link between UAV $i$ and user $j$ is lower bounded by
$C \ge \frac{1}{2}{\log _2}\left( {1 + \frac{e}{{2\pi }}{{\left( {\frac{{\xi {P_i}{h_{ij}}}}{{{\sigma _w}}}} \right)}^2}} \right)$ \cite{Wang2013Tight},
where $\sigma_w$ is the standard deviation of the additive white Gaussian noise, $\xi $ is the illumination target, and $P_i$ is the optical power of UAV $i$.
Note that LED-enabled UAVs are more likely to be deployed in dim scenarios (e.g. at night time) \cite{Deng2018twinkle}, and, hence, the effect of ambient light can be modeled as Gaussian noise.
We assume that each user has a data rate constraint ${C_\textrm{th}}$. Therefore, the required minimum optical intensity of UAV $i$ associated with user $j$ can be expressed as ${P_{{i_j},\min }} = \frac{{{\sigma _w}\sqrt {\frac{{2\pi }}{e}\left( {{2^{2{C_\textrm{th}}}} - 1} \right)} }}{{\xi {h_{ij}}}}$.
Since VLC involves both communications and illumination, the illumination constraint should also be taken into consideration.
The illuminance implies the brightness of the illuminated surface, which is proportional to
${\eta _j} = \xi {P_i}{h_{ij}} $ \cite{DinEnergyefficient}.
From the top view in Fig. \ref{scenario}, we can observe that the UAV can be carefully configured to reduce the distance between UAVs and users and, thus reducing the consumed transmit power. Therefore, the associated optimization problem is formulated as follows:
\vspace{-0.2cm}
\begin{align}\label{original_problem}
\setlength{\abovedisplayskip}{-20 pt}
\setlength{\belowdisplayskip}{-20 pt}
& {\mathop {\min }\limits_{{x_i},{y_i},{{\cal U}_i}} \sum\limits_{i = 1}^{K} {{P_i}} }\\
&\rm{s.\;t.}\scalebox{1}{$\;\xi {P_i}{h_{ij}} \ge {\rm{ }}{\eta _{{\rm{th}}}},\;\forall i \in \mathcal{N},\;j \in {\mathcal{U}_i}$}\tag{\theequation a}\\
&\scalebox{1}{$\;\;\;\;\;\;\;{P_i} \ge {P_{{i_j},\min }},\;\forall i \in \mathcal{N},\;j \in {\mathcal{U}_i}.$} \tag{\theequation b}
\end{align}
Here, $\mathcal{U}_i$ is the set of users that are serviced by UAV $i$ and $i_j$ is the index of UAV $i$ that services user $j$. In addition, $\mathcal{N}$ is the set of UAVs and $\eta\rm{_{th}}$ is the illumination threshold of the receiver.
The goal of problem (\ref{original_problem}) is to find the optimal location of each UAV $i$ $({x_i},{y_i},z_u)$ and cell association $\mathcal{U}_i$ that can minimize the total power consumption of UAVs.
However, solving (\ref{original_problem}) is challenging due to the mutual dependence between ${x_i},{y_i}$, and $\mathcal{U}_i$. Therefore, further manipulations are required to find the solution of (\ref{original_problem}).

\section{UAV Location and Cell Association Optimization}
From (\ref{channel_model}), we can observe that the channel gain decreases with distance.
This means that a UAV can successfully satisfy all of the users' requirements once the requirement of the farthest user is satisfied.
We define the farthest user as the user that has a maximum distance to its associated UAV.
Since VLC can achieve precise localization with low complexity \cite{PathakVisible}, it is reasonable to assume that the locations of all users are known.
The location of the farthest user as well as the distance between the farthest user and its associated UAV depends on the location of the UAV and the cell association.
This means that problem (\ref{original_problem}) consists of a UAV location optimization problem and a cell association problem.
We first divide the original problem into two sub-problems, i.e. UAV location optimization and cell association optimization. Then, randomized incremental construction based algorithm and a greedy strategy are used to solve these two problems, respectively. Although this proposed approach is suboptimal due to the problem division, it is efficient and avoids using exhaustive search, which requires prohibitively high complexity.
\vspace{-0.2cm}
\subsection{UAV Location Optimization}
Here, we assume that the cell-association is fixed, which implies that the users serviced by each UAV are known.
Under this assumption, the optimization problem for each UAV becomes independent.
In addition, a UAV needs to only consider its associated farthest user, as once the requirement of the farthest user is satisfied, all the other users' requirements will be satisfied.
Based on these observations, the optimization problem for each  UAV $i$ can be expressed as:
\vspace{-0.2cm}
\begin{align}\label{sub-problem1}
\setlength{\abovedisplayskip}{-20 pt}
\setlength{\belowdisplayskip}{-20 pt}
&{\mathop {\min }\limits_{{x_i},{y_i}} {P_i}}\\
&\rm{s.\;t.}\scalebox{1}{${\;\;\;\xi {P_i}{h_{ij_i^{\rm{*}}}} \ge {\eta\rm{ _{th}}},}$}\tag{\theequation a}\\
&\scalebox{1}{${\;\;\;\;\;\;\;\;{P_i} \ge {P_{{i_{j_i^*}},\min }},\;}$} \tag{\theequation b}
\end{align}
where $j_i^*$ denotes the farthest user from UAV $i$.
Note that the farthest user will dynamically change with the position of the UAV.
The optimal position of each UAV can be determined based on the following theorem.

\noindent\textbf{Theorem 1.} For a fixed cell association, a unique optimal position for UAV $i$ exists at the center of the smallest enclosing disk that covers all of the users in the illuminated region of UAV $i$.

\vspace{-0.2cm}
\begin{proof}
Substituting (\ref{channel_model}) into (\ref{sub-problem1}a), we obtain ${P_i} \ge V{\left( {{d_{ij_i^*}}} \right)^{m + 3}}$, where $V = \frac{{2\pi {\eta \rm{_{th}}}}}{{\left( {m + 1} \right)Ag\left( {{\psi _{ij_i^*}}} \right){z_u^{m + 1}}\xi }}$. Similarly, for (\ref{sub-problem1}b), we obtain ${{P_i} \ge \frac{M}{N}{{\left( {{d_{ij_i^*}}} \right)}^{m + 3}}}$, where $M = {\left( {2\pi } \right)^{\frac{3}{2}}}{\sigma _w}\sqrt {\frac{{{2^{2{C_\textrm{th}}}} - 1}}{e}} $, $N = \xi \left( {m + 1} \right)Ag\left( \psi_{ij_i^*}  \right){z_u^{m + 1}}$. Therefore, (\ref{sub-problem1}) is equivalent to:
\vspace{-0.25cm}
\small
\begin{equation}\label{transformation_equation}
\begin{array}{*{20}{l}}
{\mathop {\arg \min }\limits_{{x_i},{y_i}} {P_i} = \mathop {\arg \min }\limits_{{x_i},{y_i}} \;\max \left[ {V{{\left( {{d_{ij_i^*}}} \right)}^{m + 3}},\frac{M}{N}{{\left( {{d_{ij_i^*}}} \right)}^{m + 3}}} \right]}\\
{\begin{array}{*{20}{l}}
{\;\;\;\;\;\;\;\;\;\;\;\;\;\;\; = \mathop {\arg \min }\limits_{{x_i},{y_i}} {d_{ij_i^*}}}\\
{\;\;\;\;\;\;\;\;\;\;\;\;\;\;\; = \mathop {\arg \min }\limits_{{x_i},{y_i}} {r_{ij_i^*}},}
\end{array}}
\end{array}
\end{equation}
\normalsize
where $\max \left[ { \cdot , \cdot } \right]$ outputs the maximum of the inputs and
${r_{ij_i^*}}$ is the horizontal distance between UAV $i$ and user ${j_i^*}$ in 2D $x-y$ plane defined as ${r_{ij_i^*}} = \sqrt {d_{ij_i^*}^2 - {z_u^2}} $. When the system parameters are given, variables $M$, $N$, and $V$ are constant.
Moreover, variable ${r_{ij_i^*}}$ involves the constraint ${r_{ij_i^*}} \ge {r_{i{j_i}}}$ for ${j_i} \in {\mathcal{U}_i}$, where $j_i$ denotes the $j$-th user serviced by UAV $i$.
\emph{Therefore, the goal of problem (\ref{transformation_equation}) is to find the optimal position of the center of the enclosing disk that minimizes the radius ${r_{ij_i^*}}$}.

Next, we show that the optimal location of UAV $i$ is unique. Suppose there are two smallest enclosing disks ${D_1}$, ${D_2}$ with radius $r$. All the users are in ${D_1} \cap {D_2}$, which are further contained in a disk with radius $\sqrt {{r^2} - {a^2}} $, where $a$ is half the distance between the two disk centers. Therefore, $a$ has to equal to 0, i.e. ${D_1}$ and ${D_2}$ are the same, or a disk that contains all users with smaller disk exists. This ends the proof.
\end{proof}

\vspace{-0.3cm}
From Theorem 1, given the locations of users, the smallest enclosing disk can be obtained by randomized incremental construction \cite{Welzl1991Smallest}. A graphical representation of this method is given in Fig. 2, in which $U_i$ is the number of users serviced by UAV $i$, and $D_{j-1}$ is the disk that covers all the previous users from 1 to $j-1$.
With the obtained smallest disk $D_{U_i}$ as well as its radius $r_i$ from Fig. 2, the minimum power of UAV $i$ can be derived as
${P_{i,{\rm{min}}}} = \max \left[ {\frac{M}{N}\sqrt {r_i^2 + {z_u^2}} ,V\sqrt {r_i^2 + {z_u^2}} } \right].$

\normalsize
\begin{figure}[tp]
\begin{center}
\includegraphics[width=0.43\textwidth]{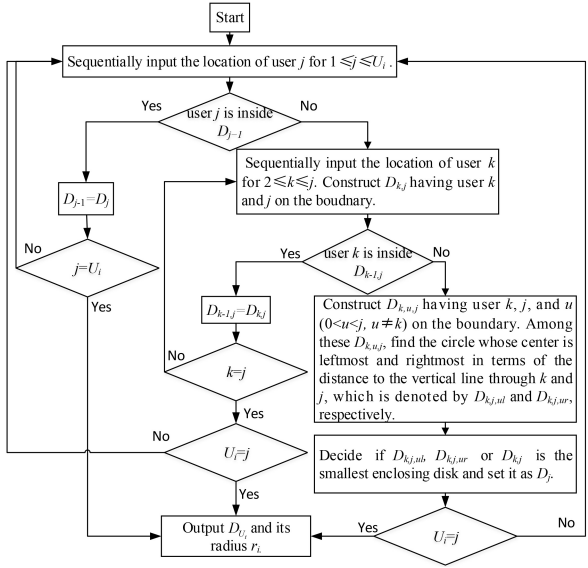}
\caption{Graphical representation of the proposed UAV location optimization approach.}
\label{proposed_method}
\end{center}
\end{figure}
\vspace{-0.4cm}
\subsection{Cell Association Optimization}
From Theorem 1, we can observe that the required power of a UAV is determined by its distance to the farthest user ${d_{ij_i^*}}$.  Since this distance not only depends on the position of the UAV but also on the locations of the users, next, we further reduce the power consumption by cell association. The optimization problem is formulated as:
\vspace{-0.2cm}
\begin{align}\label{sub_problem2}
\setlength{\abovedisplayskip}{-20 pt}
\setlength{\belowdisplayskip}{-20 pt}
&{\mathop {\min }\limits_{{\mathcal{U}_i}} \sum\limits_{i = 1}^{K} {{P_i}} }\\
&\;\rm{s.\;t.}\scalebox{1}{$\;\;\;\;\xi {P_i}{h_{ij}} \ge {\rm{ }}{\eta _{{\rm{th}}}},\;\forall i \in \mathcal{N},\;j \in {{\mathcal{U}}_{i}}$}\tag{\theequation a}\\
&\scalebox{1}{$\;\;\;\;\;\;\;\;\;\;\;{{P_i} \ge {P_{{i_j},\min }},\;\forall i \in \mathcal{N},\;j \in {{\mathcal{U}}_{i}}}$} \tag{\theequation b}
\end{align}
Similarly, from (\ref{sub_problem2}), we can find:
\small
\begin{equation}\label{sub_problem2_transferred}
\begin{array}{l}
\mathop {\arg \min }\limits_{{\mathcal{U}_i}} \sum\limits_{i = 1}^{K} {{P_i}}  = \mathop {\arg \min }\limits_{{\mathcal{U}_i}} \max \left[ {V\sum\limits_{i = 1}^{K} {d_{ij_i^*}^{m + 3}} ,\frac{M}{N}\sum\limits_{i = 1}^{{K}} {d_{ij_i^*}^{m + 3}} } \right]\\
\;\;\;\;\;\;\;\;\;\;\;\;\;\;\;\;\;\;\;\;\;\;\;\;\; = \mathop {\arg \min }\limits_{{\mathcal{U}_i}} \sum\limits_{i = 1}^{K} {d_{ij_i^*}^{m + 3}}.
\end{array}
\end{equation}
\normalsize
From (\ref{sub_problem2_transferred}), we can observe that the problem becomes a classical clustering problem \cite{AltMinimumCost}, i.e. minimizing the cost $d_{ij_i^*}^{m + 3}$ of each disk while covering all the users.
It has been proved that this problem is NP-hard when the power of the exponent cost function is more than 1 \cite{AltMinimumCost}. Therefore,  a suboptimal algorithm is proposed for the cell association which is based on a greedy strategy \cite{AltMinimumCost}.
It works as follows: 1) Start with $K$ disks with center at each UAV and zero radius. 2) Sequentially input the location of each user and find the disks among all $K$ disks that requires least cost function growth. 3) Repeat until all the users are covered by a disk and an eligible cell association is thus generated.

Since sequentially solving (\ref{sub-problem1}) and (\ref{sub_problem2}) cannot guarantee that the obtained solution is globally optimal, multiple iterations are required to further reduce the power consumption. Explicitly, with the obtained suboptimal cell association, the updated UAV locations can be obtained from (3). Then, the cell association can be updated again utilizing the updated UAV locations. The iterations end when the algorithm is not able to further reduce the power consumption. In practice, the algorithm can run in centralized fashion by a central UAV.

\section{Numerical Results}
We consider a square area, which is further divided into four sub-areas. In each sub-area, a UAV is deployed to provide communication and illumination.
The users are uniformly distributed in the considered area.
In this work, we use LEDs with a relatively wide semi-angle to illuminate larger ground areas at a fixed height. Unless otherwise specified, all the system parameters are listed in Table I. Note that since parameters such as the incidence and irradiance angles of a VLC link are determined by the positions of the UAVs and users, they are not listed in the Table I. All statistical results are averaged over a large number of independent runs.

In the initial state, the UAV is in the center of each square sub-area, and the users in each sub-area are associated with the UAV in the sub-area.
When the target area is fixed, it is well known that the center is the optimal location for the UAV \cite{KomineFundamental}, and this serves as the baseline scheme in our simulations. In particular, in static algorithm 1 (SA1), we assume the positions of the users are known at the UAV, and the power of each UAV is determined by the distance from the UAV at the center of the sub-area to the location of its corresponding farthest user.
Static algorithm 2 (SA2) is the worst case of SA1, i.e. always assuming the farthest user is on the farthest geometry point of the servicing area to the center.
In addition, we also simulate a deployment with UAV location optimization only (UAVOO), to clearly show the performance gains obtained by the UAV location optimization and the cell association optimization.
\begin{table}[tp]
\small
\centering  
\caption{System Parameters.}
\vspace{-0.2cm}
\begin{tabular}{lccc}
\hline
Name of Parameters &Values \\ \hline  
Size of the considered area/sub-area& 10 $\times$ 10 $\rm{m^2}$/ 5 $\times$ 5 $\rm{m^2}$\\
Refractive index $n_r$ & 1.5\\
Semi-angle at half power $\Phi _{1/2}$ &$60^\circ $ \\
Receiver FOV’s semi-angle $\Psi_c$ &$60^\circ $ \\
Detect area of photodiodes, $A$ & 1 $\rm{cm^2}$\\
height of UAV $z_u$ & 8 $\rm{m}$\\
\hline
\end{tabular}
\label{table}
\end{table}
\normalsize

\begin{figure}[tp]
\begin{center}
\includegraphics[width=0.45\textwidth]{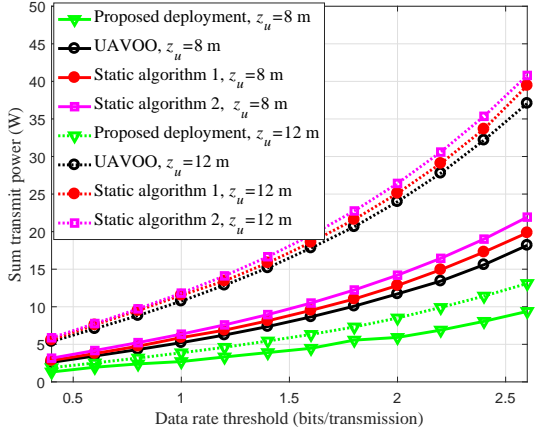}
\vspace{-0.2cm}
\caption{{The required sum transmit power of four UAVs at different heights versus the data rate threshold of users.}}
\label{datarate_power}
\end{center}
\end{figure}
In Fig. \ref{datarate_power}, we analyze the required sum transmit power for different data rates and illumination thresholds.
As we can observe from  Fig. \ref{datarate_power}, the proposed deployment will always yield the minimum required power when compared to the other three schemes.
For instance, when the required data rate threshold is 2 bits/transmission, the proposed deployment consumes approximately $53.8\%$, $57.14 \%$, and $60 \%$ less power than UAVOO, SA1, and SA2, respectively, for a UAV height of 8 m.
In addition, the proposed approach can even achieve additional performance gains for UAVs that fly at a height of 12 m. This is due to the fact that a higher altitude implies a larger ground can be covered by a UAV. By properly optimizing cell association process, fewer UAVs can be deployed and, thus, the performance gains become more pronounced.

\begin{figure}[tp]
\begin{center}
\includegraphics[width=0.45\textwidth]{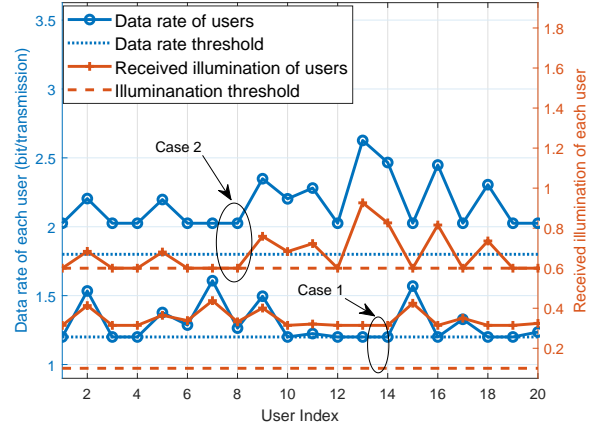}
\vspace{-0.2cm}
\caption{{The data rate and received illumination level of each user. Two cases are considered: In case 1, $C_{\rm th}=1.2$, $\eta_{\rm th}=0.1$; In case 2, $C_{\rm th}=1.8$, $\eta_{\rm th}=0.6$.}}
\label{eachuser}
\end{center}
\end{figure}

Fig. \ref{eachuser} shows the data rate and received illumination level of each user. Two typical cases are considered. In particular, in case 1, the data rate constraint requires a higher sum transmit power than the illumination constraint. Therefore, the data rate of certain users are slightly greater than or exactly equal to the data rate threshold while the received illumination levels of all users are above the illumination threshold.  In contrast,  the sum transmit power required by the illumination constraint is higher than the one required by the data rate constraint and, hence, we can observe that the illumination of certain users is slightly greater than or exactly equal to the illumination threshold while the data rates of all users are higher than the data rate threshold. However, in both cases, the data rate and the illumination requirements of each user can be satisfied.
\vspace{-0.2cm}
\section{Conclusion}
In this paper, we have studied the problem of optimal deployment of VLC-based UAVs that can provide both illumination and communication.
We have formulated the problem as a power minimization problem under illumination and communication constraints.
We have then proposed a two-step approach to solve the problem. We have shown that the problem can be separated as a smallest enclosing disk problem and a min-size clustering problem. To solve these problems, we have applied randomized incremental construction to obtain the optimal UAV locations and a greedy method to obtain a suboptimal cell association. Numerical results show that the proposed approach can yield more than $38.5\%$ improvements in power efficiency.
\vspace{-0.5cm}
\bibliographystyle{IEEEbib}
\bibliography{young}

\end{document}